\documentclass[runningheads]{llncs}
\usepackage{graphicx}
\usepackage{multirow}
\usepackage[hyphens]{url}
\usepackage{hyperref}
\usepackage[hyphenbreaks]{breakurl}
\usepackage[disable]{todonotes} 

\begin{document}
\title{Self-Driving Cars and Driver Alertness\
}
%
%
\author{Nguyen H Tran  \and
Abhaya C Nayak}
\authorrunning{N. Tran and A. Nayak}
%
\institute{Macquarie University, Sydney, Australia\\
\email{Nguyen-Huong.Tran@students.mq.edu.au | Abhaya.Nayak@mq.edu.au}
}
\maketitle              
\begin{abstract}
\todo{Abstract {\sc has been rewritten}}
Recent years have seen growing interest in the development of self-driving vehicles that promise (or threaten) to replace human drivers with intelligent software. However, current self-driving cars still require human supervision and prompt takeover of control when necessary. Poor alertness while controlling self-driving cars could hinder the drivers’ ability to intervene during unpredictable situations, thus increasing the risk of avoidable accidents. In this paper we examine the key factors that contribute to drivers’ poor alertness, and the potential solutions that have been proposed to address them. Based on this examination we make some recommendations for various stakeholders, such as researchers, drivers, industry and policy makers.
\end{abstract}

Self-driving cars, or autonomous cars, refer to cars that rely on sensory and image data of the surrounding environment to navigate. By replacing human drivers with intelligent software, they promise to make road travel safer \cite{DBLP:journals/access/YurtseverLCT20}. 
To classify self-driving cars with varying levels of automation, governments and the automotive industry often adopt the six levels of automation proposed by the Society of Automotive Engineers (SAE) \cite{SAE}. While Level 0 represents no automation whatsoever, Level 1 and 2 self-driving cars include basic lane-centering or adaptive cruise control to support the drivers in limited scenarios. As autonomous control increases, Level 3 through 5 cars include automated driving features that can replace human drivers in most scenarios.

Currently, the technologies for Level 5 fully automated self-driving cars are still evolving. Hence, most self-driving cars on the roads fall under Levels 3 and 4. In such highly automated cars, when automated driving features are engaged, human drivers are technically not driving. Nevertheless, they are expected to constantly supervise and take over immediate control in case technology fails, or other difficult circumstances arises unexpectedly \cite{Fav}. 

Furthermore, a new set of risks emerges when a human and a machine ``share the wheel'' on Level 3 and 4 self-driving cars. For example, the official investigation in 2020 by the National Transport Safety Board concluded that \emph{the Tesla Autopilot system’s limitations}, \emph{the driver’s over-reliance on the Autopilot} and \emph{the driver’s distraction} led to the 2018 fatal crash of a Tesla Model X \cite{NationalTransport}. This tragic accident corroborates the perception that over-reliance on self-driving cars combined with poor alertness on part of the driver could result in avoidable loss of lives.
A survey in 2019 confirmed that nearly three in four people in the United States are afraid of fully self-driving cars and 73\% of Americans expressed a lack of trust in the technology’s safety \cite{MohnTanya}. This strongly suggests that there is a need for car manufacturers to improve safety records in order to effectively promote the adoption of Level 3 and 4 self-driving cars. 

The safety of self-driving experience involves both technological factors as well as human behaviour.
Although technologies are often blamed for it, poor alertness on part of the driver has demonstrably played a role in virtually all crashes \cite{InsuranceInst20}. 
This is due to a number of factors including fatigue, distraction while supervising self-driving cars and attempts at taking over control amidst dangerous road traffic \cite{DBLP:conf/hci/SalisCMAKMWC20}.

It is important to gain a better understanding of the reasons behind drivers' poor alertness in order to address it.
We do that in the following Section~\ref{sec:understand}.
This understanding puts in context the remedies that have been proposed; these are outlined in Section~\ref{sec:remedy}.
In particular we note that some of these remedies focus on driver behaviour modification, and others on technological innovation.
In the subsequent Section~\ref{sec:discuss} we analyse the proposed remedies vis-a-vis the factors that contribute to poor alertness.
This helps us to identify the weaknesses in the existing approaches to redress poor alertness issues, and recommend further remedial measures that different stakeholders need to undertake.
Finally, in the concluding Section~\ref{sec:conclude} we outline the future outlook after a brief summary.

\section{Behind the Poor Alertness}\label{sec:understand}
%
As we mentioned above, while cars of the future might be fully autonomous, existing Level 3 or 4 self-driving cars still require human drivers to constantly supervise and remain ready to immediately take back vehicle control if necessary. However, since drivers are unlikely to be able predict when such handover situations would arise, poor alertness could hinder their ability to respond to emergencies. In order to promote better safety for self-driving cars, various studies have been conducted to investigate common factors that contribute to drivers' poor alertness. In this section, we explore three of the main contributing factors in order that a better understanding is gained of \emph{why human drivers become negligent on self-driving cars}. This understanding is expected to help design remedies to amend the situation.

\subsection*{Passive Fatigue}
Vehicle automation leads to more monotonous driving experience on part of the human drivers.\footnote{Those who have driven both the older manual gear shift cars with clutches and the relatively newer automatic cars would probably testify to this from personal experience.} During monotonous driving, drivers could easily become bored, disengaged and less attentive to traffic conditions \cite{MayJenniferF2009DfTi}. Such an impaired mental state of reduced alertness is often known as passive fatigue -- fatigue caused by a lack of active engagement. Simulated driving experiments have confirmed that longer periods of monitoring self-driving cars see drivers exhibiting slower braking and poorer steering responses during take-over, thus increasing the probability of a crash \cite{SaxbyDyaniJ2013AaPF}. Another study has shown that supervising self-driving cars trigger passive fatigue within 25 minutes and impeded take-over performance after 50 minutes \cite{JaroschOliver2019Eonr}. In other words, passive fatigue results in deteriorated reaction time and impaired perception, consequently contributing to poorer take-over from automated driving.

\subsection*{Distraction}
Apart from passive fatigue, automated driving also leads to prolonged boredom as evidenced by recent studies \cite{CunninghamMitchellL2018Ddai}. Being bored encourages drivers to engage in non-driving-related activities \cite{DBLP:journals/hf/CarstenLBJM12}, resulting in diminished ability to supervise and respond to take-over situations. Similarly, the higher the level of automation, the more likely the drivers would divert their visual attention away from the road ahead \cite{LlanerasRobertE2013HFIA}. A simulated driving study revealed that the presence of in-car distractions such as video and mobile phones made drivers slower at reacting to red traffic lights \cite{DBLP:conf/automotiveUI/ArkonacBSB19}. Evidently, the propensity for drivers to become less alert could heighten the risk of avoidable accidents.

\subsection*{Over-reliance on automation}
While automation on vehicles leads to boredom, perhaps somewhat paradoxically, evidence shows that, over time, drivers tend to trust automatic cars, and that trust  in turn leads to over-reliance on automation \cite{DBLP:journals/hf/LeeS04}. More trust in self-driving cars tends to degrade the driver's ability to regain vehicle control \cite{DixitVinayakV2016AVDA}. Additionally, if drivers rely more on automation to bring them to a safe stop, they feel justified engaging in distracting activities \cite{DBLP:journals/hf/CarstenLBJM12}. Given the small time window between accidents and safe takeovers \cite{HancockP.A2019Spit}, such findings offer valuable insights on how long-term usage of self-driving cars could hinder drivers’ alertness during an emergency. Unfortunately, the relevant evidence is not statistically significant, and the explanation about the possible relationships between over-reliance on self-driving features and drivers’ poor alertness are insufficient and inconclusive.

\section{Remedies  for Poor Alertness}\label{sec:remedy}
We have seen how the alertness of self-driving car drivers tends to diminish  as a result of passive fatigue, distraction and over-reliance on automation. The consequent crashes related to self-driving cars cast doubt on the safety of technology itself. The validity of safety concerns notwithstanding, it is crucial to recognise that the problem of drivers' poor alertness is intimately connected with human behaviour. With the right kind of support via training and technologies, human drivers could be put in a position to control self-driving cars and handle take-over scenarios more effectively. In this section we illustrate specific approaches that can be employed to empower human drivers with training and technologies to enhance safety in the driving of self-driving cars.

\subsection{Behavioural Training for Human Drivers}
There is more to driving than simply switching on the engine and turning the steering wheel. Good driving presupposes internalised (and good) knowledge of traffic laws, road safety protocols and recommended steps to handle high-risk scenarios. For autonomous cars, pertinent behavioural training prior to obtaining the driving licenses is even more crucial owing to the complexity of automated features, the inherent possibility of technical failure, and utter unpredictability of when control needs to be transferred from machine to human. Multiple studies have already explored different aspects of driver training to mitigate drivers' poor alertness on self-driving cars, as we outline below. 

\subsubsection*{Interactive Learning.}
Prior research  indicate that learning to properly interact with the self-driving cars in a risk-free environment provides a good starting point in ensuring road safety. Traditionally, non-interactive methods such as training videos and presentation slides are often used to provide basic information about self-driving capabilities. Nevertheless, researchers have explored alternative learning approaches. For example, Abdelgawad and colleagues proposed the use of driving simulators with 3D interactive displays to better familiarise new drivers with the automated driving features prior to their first ride \cite{Abdelgawad17}. Virtual Reality systems have also been suggested as an ideal training platform since they offer better opportunity to enhance drivers' interaction with self-driving cars under varying circumstances \cite{SPORTILLO2018102}.  

\subsubsection*{Takeover Scenarios.}
Interactive features in learning aside, training content plays a significant role in helping human drivers to appropriately adapt their behaviour while handling of self-driving cars in real-traffic conditions. A study in 2008 suggested that  
drivers who have been exposed to automation failures during training performed better in regaining control of self-driving cars
 than those without prior practice \cite{Bahner2008}. In a recent work, Sibi and colleagues also proposed to incorporate takeover scenarios in driver training courses for highly automated cars \cite{SibiSrinath2020BtSI}. For instance, during theoretical training, new drivers should learn about common reasons why the automated driving system would disengage and how to get ready to regain control of the vehicles to avoid accidents. Practical sessions could provide new drivers with hands-on experience related to sudden takeover requests in a low-risk environment \cite{SibiSrinath2020BtSI}. By doing so, drivers are empowered with both knowledge and practical skills to safely operate self-driving cars and regain control of the vehicles whenever necessary. 

\subsection{Technological Innovation}
While interactive learning and adequate practice might reinforce the right knowledge and positive driving habits in target drivers of self-driving cars, they cannot guarantee constant focus  of the drivers on the road ahead. So there is a need for additional mitigation strategies to address drivers' poor alertness on the real road. A viable option is to improve the design of self-driving cars through technological innovation. Various studies have proposed a wide range of novel technologies to reduce passive fatigue, re-engage distracted drivers and mitigate over-reliance on automation. Such technology solutions roughly fall under three main mitigation strategies, as described below.
 
\subsubsection*{Detection.}
To detect passive fatigue and distraction, an \emph{intelligence driver-assistance system} could be implemented to monitor drivers’ psychological state via in-vehicle cameras, sensors and wearables \cite{DBLP:conf/embc/ChangFC16}. While distraction is often detected with gaze patterns, degree of head rotation and eye movements, passive fatigue is spotted via eye-blinking duration and yawning frequency. Upon detecting drivers’ fatigue or distraction, the system would provide visual, auditory or haptic cues to re-engage drivers with traffic conditions \cite{CunninghamMitchellL2018Ddai}.  

\subsubsection*{Prevention.}
Various studies have focused on \emph{in-vehicle interaction design} to enhance drivers’ situational awareness in self-driving cars. Interactive games are suggested as a novel way to address passive fatigue by increasing the cognitive demand of the driving tasks. For example, Trivia Games via windshield display, or increased volume of music, can help drivers to regain attentional capabilities \cite{ColletChristian2019AVAW}. However, assigning a more stimulating secondary task to the driver could easily distract them from their primary supervising task \cite{DBLP:journals/sensors/LeeKJKCY20}, raising serious doubt about its validity.
To mitigate distraction, Beattie and colleagues have proposed to incorporate an in-vehicle auditory display to provide drivers with traffic alerts \cite{DBLP:conf/nordichi/BeattieBHM14}. An alternative solution is based on visualisation of traffic updates positioned near the centre of the road
 with 3D display technology and augmented reality. Such a \emph{head-up display} (HUD) could encourage distracted drivers to gaze toward the road centre, and thus remain aware of the traffic \cite{DBLP:conf/automotiveUI/WiegandMHH19}. 

\subsubsection*{Correction.}
Assuming that the drivers are already distracted or experience passive fatigue, previous studies have focussed on two main approaches to correct poor alertness. 
The first approach aims to design a more effective method to signal takeover requests. 
For example, since auditory and visual senses of distracted drivers are already impaired, vibrotactile alerts could be more useful to signal take-over requests \cite{Cohen-LazryGuy2019Dtaf}. 
In another study, Bazilinskyy and colleagues have proposed a multimodal take-over request that combines auditory, visual and vibrotactile displays for high-urgency warnings since a single mode of alert could easily fail or become overloaded \cite{BazilinskyyP2018Trih}. 
The second approach to correct driver alertness focuses on developing an automated fallback function to prevent crashes when the system cannot rely on the driver to bring the vehicle to a safe stop. 
In a traffic jam pilot to model how an automated fallback function should work, the self-driving car was programmed to first issue a request to the driver to take over during an emergency. If the driver failed to respond within the allowed timeframe, or a critical system failure occured, the self-driving cars was programmed to slow down automatically, while concurrently sending out warning signals to alert surrounding traffic participants \cite{BeckerJan2016SAaS}. 

\section{Discussion and Recommendations}\label{sec:discuss}
It is evident that human drivers on self-driving cars find it challenging to stay alert and effectively respond to emergency events due to passive fatigue, distractions and excessive trust in automation. Earlier we discussed a number of solutions  designed to inculcate expected behaviour in human drivers, and enhance their ability to handle takeover requests. In this section, we examine the limitations of existing approaches to tackle the problem of poor alertness, and provide suggestions to further enhance safety for Level 3 and 4 self-driving cars.  

It is clear that different technologies can address different factors contributing to drivers’ poor alertness on self-driving cars \cite{MayJenniferF2009DfTi}.
Table~\ref{tab1}  
below presents a summarised mapping between technological solutions and the three key factors that primarily contribute to drivers’ poor alertness on self-driving cars. 

\begin{table}
\caption{Mapping  technological solutions to factors contributing to poor alertness.}\label{tab1}
\begin{tabular}{||l||l||c|c|c||}\hline\hline
  \multirow{2}{*}{\bf Strategy} & \multirow{2}{*}{\bf Technological Solution} &  \multicolumn{3}{|l||}{\quad\quad\bf Contributing Factors} \\ \cline{3-5}
  & &  {\em Fatigue} & {\em Distraction} & {\em Overreliance}  \\ \hline\hline
 {\bf Detection} & Intelligent driver-assistance system & $\times$ & $\times$  & {} \\ \hline\cline{1-5}
\multirow{3}{*}{\bf Prevention} & Interactive games or music &  $\times$  & {} & {} \\ \cline{2-5}
& In-vehicle auditory display (alerts) &  {} & $\times$  & {} \\ \cline{2-5}
& 3D \& AR head-up display &  {} & $\times$  & {} \\ \hline\cline{1-5}
\multirow{2}{*}{\bf Correction} &Multimodal take-over request &  $\times$  & $\times$  & {} \\ \cline{2-5}
 &Auto. fallback function (crash prevn.) &  $\times$  & $\times$  & $\times$   \\ \hline
\hline\end{tabular}
\end{table}

A quick glance at Table~\ref{tab1} above shows that the majority of the existing solutions  address only one or two  factors that contribute to drivers’ poor alertness. Since a combination of passive fatigue, distractions and over-reliance on automation might affect drivers at the same time \cite{DBLP:journals/sensors/LeeKJKCY20}, these proposed solutions might not be sufficiently effective in real life. Moreover, since existing solutions have been developed solely for research purpose, detail implementation plan as to how they will coexist with human drivers and other components of self-driving cars are often incomplete. There is a need for further research aiming to develop holistic solutions that address all three contributing factors and consider practical implementation concerns, in order that car manufacturers would be encouraged to adopt and develop these novel ideas for commercialisation.

\subsection{Pre-Driving Safety}
Various studies concur that existing self-driving cars require human drivers to shift from manually controlling the cars to ``merely'' supervising the automated driving features. Hence, existing safety requirements and driver license prerequisites designed for standard vehicles might not be sufficient ,nor relevant, for self-driving cars. There is a clear need for additional measures to ensure that both the car and the human driver are ready before such cars become available to the public. 

\subsubsection*{Driver Training.}
Undergoing special training prior to obtaining the driving licenses is crucial for drivers' safety owing to the inherent possibility of technical failure in autonomous vehicles. The complexity of automated driving features highlights the need for different training strategies to support human drivers on self-driving cars. Specifically, interactive learning and hands-on experience with takeover requests result in better outcomes as compared to traditional videos and presentations. While 3D interactive displays and Virtual Reality could facilitate better interaction between human drivers and the vehicles during training, the efficacy of these training programmes in reducing crashes on the real road remains uncertain. Additionally, a standardised framework for driver training that are tailored to Level 3 and 4 self-driving cars is yet to be developed. Since trainee drivers should undergo consistent training programs, it is urgent that requisite standards for such training programs be developed.

\subsubsection*{Driver Licensing.}
Although a good knowledge of traffic laws, road safety protocols and recommended steps to handle high-risk scenarios are sufficient to obtain driver licenses for standard cars, these might be inadequate to ensure safe operations of self-driving cars. Unfortunately, existing laws still follow non-autonomous driving policies \cite{DBLP:journals/see/Ryan20}, and the standardised driving tests are an insufficient measure of a person's ability to control self-driving cars. To enhance safety, drivers should undergo  trainings that additionally cover knowledge and skills related to self-driving cars.
Furthermore, strict requirements related to handling takeover requests and responding to system failure should be incorporated into driving tests before issuance or renewal of driving licenses for Level 3 and 4 self-driving vehicles.

\subsubsection*{Vehicle Safety Requirements.}
Apart from rigorous driver training and testing schedules, the safety of autonomous vehicles themselves should also be carefully ensured. Since self-driving cars should be as safe as non-autonomous cars, many jurisdictions have set specific restrictions to establish minimum safety standards for self-driving cars before approving sales to the general public. Although safety specifications might substantially differ across governments, there is a need to establish a common baseline for safety requirements. Specific technologies to facilitate smooth transition between automated system and human drivers such as multi-modal takeover request or in-vehicle detection of driver's fatigue should be made mandatory to adequately protect drivers and passengers on self-driving cars. 

\subsection{Ongoing Safety Requirements}
Although driver licenses and vehicle safety requirements prepare the driver for the first ride, safely navigating the road over a prolonged period of time requires ongoing efforts to reinforce positive driving habits and enhance confidence on self-driving cars. We elaborate below the  need for refresher driving lessons and feedback mechanisms for in-car safety features.  

\subsubsection*{Refresher Driving Training.}
As it is, standard driving licences often require extra checks before renewal -- specifically if the licence holder is insulin dependent, has other such disability, or is too elderly.
At the moment it is not known if additional such factors (possibly based on cognitive abilities of the licence holder) have negative impact on one's ability to supervise self-driving cars.
So there is a likely need to strengthen the requirements for renewal of licence for self-driving cars.
System updates on automated driving features, recent changes in traffic laws, or recently discovered high-risk scenarios could  form part of the refresher driving lessons in order to mitigate the risk of poor alertness. 
Furthermore, there is very little critical assessment of the implications of prolonged ``exposure'' to self-driving features.
Short of adequate data, reintroducing safety precautions and best practices on handling sudden takeover requests could strengthen good behaviour in human drivers. 

\subsubsection*{Feedback Mechanism for Safety Features.}
The proposed technological solutions show promise in addressing poor alertness.
However, they are mostly based on studies that,  owing to safety concerns, involved simulated driving or short, well controlled, test-driving  \cite{DBLP:journals/sensors/LeeKJKCY20}. 
Furthermore, prolonged real-life driving are expected to involve varying psychological complexities, raising doubt the effectiveness of these technologies. 
Hence, more on-road driving studies on human behaviour in the context of self-driving cars are required to improve safety features  addressing poor alertness. 

\subsection{Recommendations for Stakeholders}

There is a somewhat flawed illusion that during an emergency or system failure, the driver can simply take over control and bring the vehicle to a safe stop. 
\todo{originally Google is convicted??? [From Nguyen] Originally it was Google's conviction. Let's keep it as convinced. Sounds better. Thanks for the edit.}
Even Google is convinced that human and machine should not share the wheel because an inattentive driver taking over the vehicle control could be disastrous \cite{Lipson}. 
However, since the transition to Level 5 fully automated cars is still underway, for the time being drivers are expected to constantly supervise self-driving vehicles and remain ready to take over. Based on our earlier discussion, we outline below some key recommendations for different stakeholders, namely, car manufacturers, policy makers, consumers and researchers. 

\subsubsection*{Car manufacturers.}
To promote the adoption of Level 3 and 4 self-driving cars, improved safety record is the top priority. Car manufacturers should continue to invest in new technological competencies around the automated fallback function to prevent avoidable crashes. They should also collaborate with researchers, sharing real-life anonymised data about human behaviour on highly automated vehicles. This will help develop and commercialise innovative solutions for preventing, detecting and correcting drivers’ poor alertness. 
\subsubsection*{Governments and Policy Makers.}
As self-driving cars go mainstream in developed countries, automation is expected to drastically change on-road human behaviour. 
Governments should  invest in research on the long-term effect of autonomous vehicles on human’s driving behaviour, and support development of  automated fallback function to bring the vehicle to a safe stop when necessary. 

Funding for research and development aside, there is a need for development on the policy front.
A common set of standards for operating Level 3 and 4 self-driving cars should be urgently developed. 
In particular,  stricter requirements should be enforced for driver training and driving tests before issuance or renewal of driving licenses for individuals seeking to operate Level 3 or 4 self-driving cars. 
Moreover, given the increased likelihood of diminished alertness on part of the drivers, mitigating technological features should be part of the safety requirements  before a self-driving car is certifed roadworthy.  
\subsubsection*{Consumers.}
Existing self-driving cars  require human drivers to constantly supervise and remain ready to take back  vehicle control if necessary. 
Hence, potential consumers must carefully consider their supervisory responsibilities. 
They should also ensure availability of safety features that facilitate emergency takeover and crash avoidance when purchasing Level 3 or 4 self-driving cars. 
These include multimodal take-over request, intelligence driver assistance system and automated fallback function. 
They should proactively undertake adequate trainings that are tailored to self-driving cars before their first ride on the road.
\subsubsection*{Researchers.}
There is a need for further exploration and critical assessment of how long-term over-reliance on automation might impact drivers’ ability to respond to extreme  scenarios. 
Researchers are also recommended to develop holistic solutions that address passive fatigue, distraction and over-reliance on automation while proactively addressing real-world implementation concerns. 
Future research should also focus on more on-road driving studies and collaboration with industry partners to obtain pertinent data on human behaviour. 

\section{Conclusion and Future Outlook}\label{sec:conclude}
Self-driving cars are not yet sufficiently reliable and robust to handle all traffic conditions. 
Human drivers need to remain alert for unpredictable events requiring intervention and take-over -- self-driving cars cannot yet afford negligent drivers.
We explored three main factors that contribute to why drivers become negligent when operating self-driving cars. 
These include passive fatigue, distraction and over-reliance on automation. 
We also examined how behavioural training and technological innovation could endow the human drivers with the necessary knowledge and skill to ensure safe travel on self-driving cars. 

Admittedly, poor alertness on part of the drivers is a complex problem that has not been fully explored and understood. 
Although Level 3 and 4 self-driving cars are going mainstream in developed countries,  relevant updated driver training protocols and supporting technologies are yet to be developed to ensure road safety. 
Moreover, simulated driving is limited in its ability to  fully reflect human behaviour and the complexities of real-life driving involving highly automated cars.
Hence there is a need for further studies involving real-world scenarios in order to better understand and effectively mitigate drivers’ poor alertness.

There is also a need for better knowledge-sharing and collaboration among relevant stakeholders. 
For instance, manufacturers of self-driving car  need to provide researchers with relevant real-life, anonymised data on human behaviour, while researchers could be more proactive in incorporating industry concern so that newly developed technologies can be more acceptable to the manufacturers. 
Better conversations between consumers, car manufacturers and governments should also be encouraged in order to develop relevant policy instruments involving driver training and licensing as well as safety approval for self-driving cars. 

Other very pertinent issues, such as building of novel infrastructure by public authorities, and amendment of legal instruments to determine accident liability are outside the scope of this paper, and we will address them in our future work.

%
%


\end{document}